\documentclass[aps,prb,reprint,groupedaddress,footinbib,showpacs,citeautoscript]{revtex4-1}

\usepackage{graphicx}
\usepackage{amsmath}
\usepackage{amssymb}
\usepackage{bm}
\usepackage{color}
\usepackage{array}
\usepackage{epstopdf}

\newcolumntype {s}[1]{@{\hspace{#1}}} 

\graphicspath{{.}{./EPS/}}

\newcommand{\ket}[1]{\left | \, #1 \right \rangle}
\newcommand{\bra}[1]{\left \langle #1 \, \right |}

\newcommand* {\ee}{\mathrm{e}}

\newcommand*{\vek}[1]{{\bm{\mathrm{#1}}}}

\newcommand*{\kk}{{\bm{\mathrm{k}}}}

\DeclareMathSymbol{\myRe}{\mathord}{symbols}{"3C}
\renewcommand{\Re}{\myRe\mathrm{e}\,}
\DeclareMathSymbol{\myIm}{\mathord}{symbols}{"3D}

\begin{document}

\title{Magneto-tunneling spectroscopy of chiral two-dimensional electron systems}

\author{L. Pratley}
\email{luke.pratley@gmail.com}
\affiliation{School of Chemical and Physical Sciences and MacDiarmid
Institute for Advanced Materials and Nanotechnology, Victoria
University of Wellington, PO Box 600, Wellington 6140, New Zealand}

\author{U. Z\"ulicke}
\email{uli.zuelicke@vuw.ac.nz}
\affiliation{School of Chemical and Physical Sciences and MacDiarmid
Institute for Advanced Materials and Nanotechnology, Victoria
University of Wellington, PO Box 600, Wellington 6140, New Zealand}

\date{\today}

\begin{abstract}
We present a theoretical study of momentum-resolved tunneling
between parallel two-dimensional conductors whose charge carriers
have a (pseudo-)spin-1/2 degree of freedom that is strongly coupled to
their linear orbital momentum. Specific examples are single and bilayer
graphene as well as single-layer molybdenum disulphide. Resonant
behavior of the differential tunneling conductance exhibited as a function
of an in-plane magnetic field and bias voltage is found to be strongly
affected by the (pseudo-)spin structure of the tunneling matrix. We
discuss ramifications for the direct measurement of electronic properties
such as Fermi surfaces and the dispersion curves. Furthermore, using a
graphene double-layer structure as an example, we show how
magneto-tunneling transport can be used to measure the pseudo-spin
structure of tunnel matrix elements, thus enabling electronic characterization
of the barrier material.
\end{abstract}

\pacs{73.40.Gk,     
          73.22.Dj,      
          72.80.Vp     
}

\maketitle

\section{Introduction}

Tunneling spectroscopy is a powerful tool to probe the electronic
structure of materials~\cite{wol85}. Since the advent of microelectronic
fabrication techniques that enabled the creation of low-dimensional
electron systems, momentum-resolved tunneling transport between
parallel two-dimensional (2D) quantum wells~\cite{smo89,smo89a,eis91,
hay91,gen91,sim93,pat96,has03}, quantum wires~\cite{eug94,wan94,aus02,bie05}
and even quantum dots~\cite{vdo00} has been used extensively to measure
electronic dispersion relations~\cite{rai95,lyo06,bie06} and the effect of
interactions~\cite{aus05,jom09}. In these systems, the requirement of
simultaneous energy and momentum conservation for tunneling through
an extended barrier leads to resonances in the tunneling conductance as
the applied bias and the magnetic field parallel to the barrier are
varied~\cite{zue02}. For charge carriers subject to spin-orbit coupling,
magneto-tunneling transport has been proposed as a means to measure
the spin splitting~\cite{rai03,roz08} and to generate spin-polarized
currents~\cite{gov02,rai04,per12}.

\begin{figure}[b]
\includegraphics[width= 0.7\columnwidth]{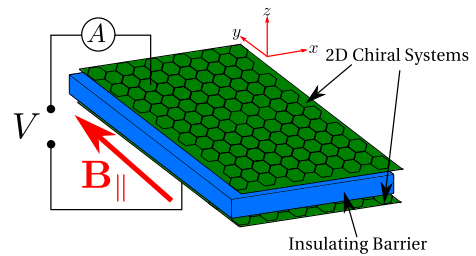}
\caption{\label{fig:diagram}
Schematics of the vertical tunneling structure considered in this work.
Two parallel chiral two-dimensional electron systems are separated by
a uniform barrier. A magnetic field applied parallel to the barrier is used
to tune resonances in the tunneling conductance that arise from the
requirement of simultaneous energy and momentum conservation.}
\end{figure}

The recent fabrication~\cite{brit12,brit12a,geo12,brit13,gei13,myo13}
of vertical field-effect transistor structures consisting of two parallel
single layers of graphene separated by an insulating barrier made of
2D crystals with large band gap opens up a new possibility to study
magneto-tunneling transport of graphene's chiral Dirac-fermion-like
charge carriers~\cite{cas09}. Unlike the real spin of electrons that is
normally conserved for tunneling through non-magnetic barriers, the
sublattice-related pseudo-spin degree of freedom of graphene electrons
can be affected by morphological details of the vertical heterostructure.
We present a systematic theoretical study of the rich variety of
pseudo-spin-dependent magneto-tunneling phenomena in vertically
separated chiral 2D electron systems. See Fig.~\ref{fig:diagram} for
an illustration of the envisioned sample geometry. Resonances in the
tunneling conductance are shown to depend sensitively on the properties
of the tunneling barrier and on whether the two parallel 2D systems are
doped with the same or opposite type of charge carriers. Our work is
complementary to previous studies~\cite{fee12,bal12,vas13} that
considered resonant behavior as a function of bias in zero magnetic field.

This article is organized as follows. We begin with a description of the
theoretical method in Sec.~\ref{sec:theory}. Results obtained for the
linear (i.e., zero-bias) magneto-tunneling conductance between various
parallel 2D chiral systems are presented in Sec.~\ref{sec:linG}. Features
arising due to a finite bias are discussed in Sec.~\ref{sec:finBiasG}.
The effect of a strong perpendicular magnetic field on tunneling between
chiral 2D systems is considered in Sec.~\ref{sec:LandQuant}. Using a
graphene double-layer system as example, we show in Sec.~\ref{sec:ExtractT}
how pseudo-spin-dependent tunnel matrix elements can be extracted
from parametric dependencies of the linear tunneling conductance.
Section~\ref{sec:concl} contains concluding remarks with a discussion
of experimental requirements for verifying our results. Certain technical
details are given in Appendices.

\section{Theoretical description of magneto-tunneling transport
\label{sec:theory}}

Heterostructures consisting of two tunnel-coupled chiral 2D electron
systems are described by a Hamiltonian of the form~\cite{vas13}
\begin{equation}
H=\begin{pmatrix}
{\mathcal H}_1 & {\mathcal T} \\ {\mathcal T}^\dagger &
{\mathcal H}_2 \end{pmatrix} \quad ,
\end{equation}
where the ${\mathcal H}_{1,2}$ are single-particle Hamiltonians
acting in the sublattice-related pseudo-spin-1/2 space for electrons
in each individual system,~\footnote{We neglect all electron-electron
interactions in this work.} and ${\mathcal T}$ is the $2\times2$
transition matrix that encodes the tunnel coupling between
pseudo-spin states from the two systems. Performing a standard
calculation~\cite{bru04} using linear-response theory for the
weak-tunneling limit yields the current-voltage ($I$--$V$)
characteristics for tunneling as
\begin{eqnarray}\label{eq:tuncurr}
&& I(V) = \frac{e}{\hbar} \sum_{\alpha\beta} \int^\infty_{-\infty}
\frac{d \varepsilon}{2\pi} \, \left[ n_{\text{F}}(\varepsilon - eV) -
n_{\text{F}}(\varepsilon) \right] \nonumber \\ && \hspace{1cm}
\times \, \mathcal{A}^{(1)}_\alpha(\varepsilon)\,
\mathcal{A}^{(2)}_\beta(\varepsilon - eV) \left|\bra{\psi^{(1)}_\alpha}
{\mathcal T}\ket{\psi^{(2)}_\beta} \right|^2  .
\end{eqnarray}
The summation index $\alpha$ ($\beta$) runs over the set of quantum
numbers for single-particle eigenstates in system 1 (2) and, thus,
generally comprises parts related to linear orbital motion,
sublattice-related pseudo-spin, real-spin and valley degrees of freedom.
$\mathcal{A}^{(m)}_\alpha(\varepsilon)$ denotes the spectral function
for single-particle excitations with quantum number(s) $\alpha$ in
system $m$ at energy $\varepsilon$, $n_{\text{F}}(\varepsilon)$ is
the Fermi-Dirac distribution function, and $\big|\psi^{(m)}_\alpha
\big\rangle$ is a single-particle eigenstate in system $m$. From
the $I$--$V$ characteristics (\ref{eq:tuncurr}), the differential conductance
\begin{equation}
G(V) \equiv \left. \frac{\partial I(V')}{\partial V'} \right|_{V' = V}
\end{equation}
can be derived. In the small-bias limit, the tunneling current (\ref{eq:tuncurr})
is proportional to the bias voltage, with the linear conductance $G(0)$ as
proportionality factor. Straightforward calculation yields
\begin{eqnarray}\label{eq:tuncond}
G(0) &=& \frac{e^2}{\hbar} \sum_{\alpha\beta} \int^\infty_{-\infty}
\frac{d \varepsilon}{2\pi} \, \left(-\frac{\partial n_{\text{F}}(\varepsilon)}
{\partial\varepsilon}\right) \, \mathcal{A}^{(1)}_\alpha(\varepsilon)\,
\mathcal{A}^{(2)}_\beta(\varepsilon)
\nonumber \\ &&\hspace{3.5cm}
\times \left|\bra{\psi^{(1)}_\alpha} {\mathcal T}\ket{\psi^{(2)}_\beta}
\right|^2 . \quad
\end{eqnarray}

In a structure with a uniform extended barrier, canonical momentum
parallel to the barrier is conserved for tunneling electrons~\cite{zhe93,lyo93,rai96}.
As a result, the tunneling matrix will be diagonal in the representation of
in-plane wave vector $\kk=(k_x, k_y)$ and, thus, can be written in the
form
\begin{equation}\label{eq:kresolvedT}
{\mathcal T} = \sum_{\kk} |\kk\rangle\langle\kk| \otimes \tau_\kk
\quad .
\end{equation}
Here $\tau_\kk$ is the momentum-resolved pseudo-spin tunneling
matrix that depends on specifics of the heterostructure. Moreover,
the single-electron eigenstates in a clean 2D chiral system from
the $\gamma$ valley ($\vek{K}$ or $\vek{K'}$ in graphene)  are
generally of the form
\begin{equation}
\ket{\psi_{\gamma,\kk,\sigma}} = \ket{\kk}\otimes
\ket{\sigma}_{\gamma, \kk} \quad ,
\end{equation}
where $\ket{\sigma}_{\gamma,\kk}$ denotes the eigenstate of
pseudo-spin-1/2 projection on a $\kk$-dependent axis. Application
of an in-plane magnetic field $\vek{B}_\parallel=B_\parallel\,
\vek{\hat b}$ (where $\vek{\hat b}$ is the unit vector in
$\vek{B}_\parallel$ direction) induces a shift between canonical
momentum $\kk$ and kinetic momentum $\vek{\Pi}^{(m)}(\kk,
\vek{B}_\parallel)$ for electrons in system $m$. A convenient
choice of gauge yields~\cite{zhe93,lyo93,rai96}
\begin{equation}
\vek{\Pi}^{(m)}(\kk,\vek{B}_\parallel)=\kk + (z_m/\ell_{B_\parallel}^2)
\,\vek{\hat b}\times \vek{\hat z} \quad ,
\end{equation}
where $z_m$ is the $z$ coordinate of system $m$ and $\ell_B =
\sqrt{\hbar/|e B|}$ is the magnetic length. The in-plane
magnetic field also modifies the pseudo-spin part of the
chiral-2D-electron eigenstates in system $m$, which then read
\begin{equation}\label{eq:BparState}
\ket{\psi^{(m)}_{\gamma,\kk,\sigma}(\vek{B}_\parallel)} = \ket{\kk}
\otimes \ket{\sigma}_{\gamma, \vek{\Pi}^{(m)}(\kk,\vek{B}_\parallel)}
\quad .
\end{equation}

\begin{figure}[t]
\includegraphics[width=0.8\columnwidth]{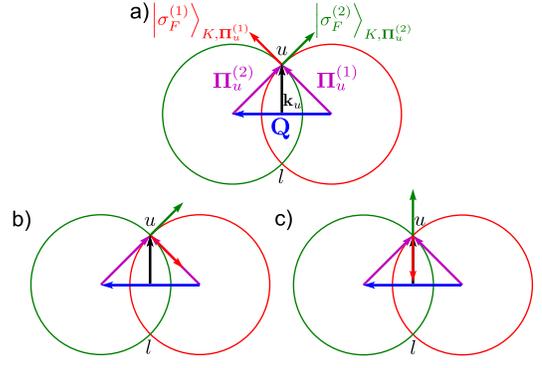}
\caption{\label{fig:fermicircles}
Visualization of constraints due to combined energy and momentum
conservation for chiral electrons. An applied in-plane magnetic field
results in a shift by $\vek{Q}$ of the Fermi circles associated with the
vertically separated 2D conductors. In the zero-bias limit, tunneling can
only occur for states at intersection points of the Fermi surfaces. For
one of the latter, the kinetic wave vectors and $\vek{K}$-valley
pseudo-spin states are indicated for the case of tunneling between
a)~two \textit{n}-doped single-layer graphene sheets,
b)~a \textit{p}-doped and an \textit{n}-doped graphene layer, and
c)~two \textit{n}-doped bilayer graphene sheets.}
\end{figure}

Inserting (\ref{eq:kresolvedT}) and (\ref{eq:BparState}) into the
expression (\ref{eq:tuncond}), using $\mathcal{A}^{(m)}_\alpha
(\varepsilon)=2\pi\delta\big(\varepsilon - \varepsilon^{(m)}_\alpha\big)$
as is applicable for noninteracting electrons with single-particle energies
$\varepsilon^{(m)}_\alpha$ in the absence of disorder, and taking the
zero-temperature limit yields the linear conductance per unit area as
\begin{eqnarray}\label{eq:conductance}
&& \frac{G(0)}{A}=\frac{g_{\text{s}} e^2}{\hbar} \sum_\gamma 
2\pi \, \rho_{\text{F}}^{(1)}\, \rho_{\text{F}}^{(2)} \, \left[ \left|
\Gamma_{\text{u}}^{(\gamma)}\right|^2 + \left|\Gamma_{\text{l}}^{(\gamma)}
\right|^2 \right] 
\nonumber \\ && \hspace{0.1cm}
\times \frac{\Theta\left( | \vek{Q} | - \left| k_{\text{F}}^{(1)} -
k_{\text{F}}^{(2)} \right| \right)\Theta \left( k_{\text{F}}^{(1)} +
k_{\text{F}}^{(2)} - | \vek{Q} | \right)}{\sqrt{\left[\left( k_{\text{F}}^{(1)}
+ k_{\text{F}}^{(2)} \right)^2 - Q^2 \right] \left[ Q^2 - \left(
k_{\text{F}}^{(1)} - k_{\text{F}}^{(2)}\right)^2 \right] } } \, .
\end{eqnarray}
Here $g_{\text{s}}=2$ is the real-spin degeneracy, $\rho_{\text{F}}^{(m)}$
the density of states at the Fermi energy in system $m$ not including
real-spin or valley degrees of freedom, $k_{\text{F}}^{(m)}$ is the
Fermi wave vector in system $m$, $\vek{Q} = [(z_2 - z_1)/
\ell_{B_\parallel}^2] \, \vek{\hat b}\times\vek{\hat z}$, and
\begin{equation}\label{eq:GammaUL}
\Gamma_{\text{u/l}}^{(\gamma)} = \, _{\gamma, \vek{\Pi}^{(1)}_{\text{u/l}}}
\!\! \bra{\sigma^{(1)}_{\text{F}}} \tau_{\kk_{\text{u/l}}}
\ket{\sigma^{(2)}_{\text{F}}}_{\!\gamma, \vek{\Pi}^{(2)}_{\text{u/l}}}
\end{equation}
are pseudo-spin tunnel matrix elements between states associated
with the two intersection points (labelled u and l, respectively) of the
two systems' shifted Fermi circles. See Fig.~\ref{fig:fermicircles} for
an illustration. The canonical and kinetic wave vectors for each of
these intersection points can be found from the conditions
\begin{subequations}
\begin{eqnarray}
\left| \vek{\Pi}^{(m)}_{\text{u/l}} \right| &=& k_{\text{F}}^{(m)}\quad , \\
\vek{\Pi}^{(1)}_{\text{u/l}} - \vek{\Pi}^{(2)}_{\text{u/l}} &=& \vek{Q}
\quad , \\
\kk_{\text{u/l}} &=& \frac{1}{2} \left( \vek{\Pi}^{(1)}_{\text{u/l}} +
\vek{\Pi}^{(2)}_{\text{u/l}} -\frac{z_1+ z_2}{\ell_{B_\parallel}^2} \,
\vek{\hat b}\times\vek{\hat z} \right) \, . \quad
\end{eqnarray}
\end{subequations}
Furthermore, the projection quantum numbers $\sigma_{\text{F}}^{(m)}$
are determined by the type of charge carriers (electrons or holes) that are
present in system $m$: $\sigma_{\text{F}}^{(m)} = +$ ($-$) if system $m$
is \textit{n}-doped (\textit{p}-doped).

To be specific, we assume from now on that the pseudo-spin tunneling
matrix $\tau_\kk\equiv \tau$ is a constant matrix and use the general
parameterization
\begin{equation}\label{eq:PauliTun}
\tau = \left( \tau_0\,\sigma_0+\tau_x\,\sigma_x+\tau_y \,\sigma_y+\tau_z
\,\sigma_z \right)/\sqrt{2}
\end{equation}
with, in general, complex numbers $\tau_j$ that encode the quantum
transfer amplitudes for various possible tunneling processes. For example,
$\tau_0$ is determined by pseudo-spin-conserving tunneling processes.
Introducing a materials-specific conductance unit
\begin{equation}
G_0 = \frac{g_{\text{s}} g_{\text{v}} e^2}{2\pi\hbar}\, \mathrm{Tr}\left[
\tau^\dagger \tau \right] \, \frac{4 \pi^2 \rho_{\text{F}}^{(1)}
\rho_{\text{F}}^{(2)}}{k_{\text{F}}^{(1)} k_{\text{F}}^{(2)}} \, A \quad ,
\end{equation}
where $g_{\text{v}}$ is the degeneracy factor associated with the valley degree
of freedom, enables us to express the magnet-tunneling conductance in a
universal form. As an example, and for future comparison, we quote the
result obtained~\cite{zhe93,rai96} for the linear tunneling conductance between
two parallel ordinary 2D electron systems with equal density and, hence, same
Fermi wave vector $k_{\text{F}}^{(1)}=k_{\text{F}}^{(2)}\equiv \bar k_{\text{F}}$:
\begin{equation}\label{eq:ord2D}
\frac{G^{(\text{ord})}(0)}{G_0} = \frac{4 \bar k_{\text{F}}^2}{Q\, \sqrt{4
\bar k_{\text{F}}^2 - Q^2}} \,\, \Theta(2 \bar k_{\text{F}} - Q) \quad .
\end{equation}

\section{Linear magneto-tunneling conductance for chiral 2D systems
\label{sec:linG}}

Results given below have been obtained through application of
Eq.~(\ref{eq:conductance}), with the pseudo-spin-dependent overlap
(\ref{eq:GammaUL}) capturing the essential differences between the
various 2D chiral systems considered here.

For electrons in a single layer of graphene, the dispersion relation is given
by~\cite{cas09} $\varepsilon_{\gamma,\kk,\sigma}^{(\text{slg})} = \sigma\, \hbar v
\, k$, and the pseudo-spin states in the $\vek{K}$ and $\vek{K'}$ valleys are
[$\theta_\kk=\arctan(k_y/k_x)$]
\begin{equation}
\ket{\sigma}_{\vek{K},\kk}^{(\text{slg})} = \frac{1}{\sqrt{2}} \left(\begin{array}{c}
\ee^{- i \frac{\theta_\kk}{2}} \\ \sigma\, \ee^{i \frac{\theta_\kk}{2}} \end{array}
\right) \, , \, \ket{\sigma}_{\vek{K'},\kk}^{(\text{slg})} = \frac{1}{\sqrt{2}} \left(
\begin{array}{c} \ee^{- i \frac{\pi - \theta_\kk}{2}} \\ \sigma\, \ee^{i \frac{\pi -
\theta_\kk}{2}}\end{array} \right) .
\end{equation}
We use these states in (\ref{eq:GammaUL}) to find the magneto-tunneling
conductance between two parallel \textit{n}-type graphene layers in terms
of $2 \bar k_{\text{F}}=k_{\text{F}}^{(2)}+k_{\text{F}}^{(1)}$ and $\Delta =
\big| k_{\text{F}}^{(2)} - k_{\text{F}}^{(1)} \big|$ as
\begin{subequations}\label{eq:SLGnn}
\begin{eqnarray}\label{eq:nnGtunn}
&& \frac{G^{(\text{slg})}_{n\leftrightarrow n}(0)}{G_0} = \frac{\Theta(Q - \Delta)
\Theta(2\bar k_{\text{F}} - Q) }{\mathrm{Tr}[\tau^\dagger \tau]}\left\{ \left[
|\tau_0|^2 + |\tau_{\perp}|^2 \frac{\Delta^2}{Q^2} \right] \right. \nonumber \\
&&\times\left. \sqrt{\frac{4\bar k_{\text{F}}^2 - Q^2}{Q^2 - \Delta^2}} + \left[
|\tau_z|^2 + |\tau_\||^2 \frac{4\bar k_{\text{F}}^2}{Q^2} \right] \sqrt{\frac{Q^2
- \Delta^2}{4 \bar k_{\text{F}}^2 - Q^2}} \right\} . \quad 
\end{eqnarray}
Here $\|$ ($\perp$) denotes the in-plane direction parallel (perpendicular) to the
magnetic field. In the case of pseudo-spin-conserving tunneling (i.e., $\tau_z=
\tau_\|=\tau_\perp=0$) and equal densities in the two layers,
Eq.~(\ref{eq:nnGtunn}) simplifies to
\begin{equation}\label{eq:nnGsp}
\frac{G^{(\text{slg})}_{n\leftrightarrow n}(0)}{G_0} = \frac{\sqrt{4\bar
k_{\text{F}}^2 - Q^2}}{Q}\,\, \Theta(2\bar k_{\text{F}} - Q)
\quad .
\end{equation}
\end{subequations}
When one of the systems is \textit{p}-type and the other \textit{n}-type, we find
\begin{subequations}\label{eq:SLGnp}
\begin{eqnarray}\label{eq:npGtunn}
&& \frac{G^{({\rm slg})}_{n\leftrightarrow p}(0)}{G_0} = \frac{\Theta(Q-\Delta)\Theta
(2\bar{k}_F-Q)}{\mathrm{Tr}[\tau^\dagger\tau]} \left\{ \left[|\tau_0|^2+|\tau_{\perp}|^2
\frac{\Delta^2}{Q^2} \right] \right. \nonumber \\ 
&&\times\left. \sqrt{\frac{Q^2-\Delta^2}{4\bar{k}_F^2-Q^2}} +\left[|\tau_z|^2 +|\tau_\||^2
\frac{4\bar{k}_F^2}{Q^2}\right]\sqrt{\frac{4\bar{k}_F^2-Q^2}{Q^2-\Delta^2}}\right\}
\end{eqnarray}
in the most general case. In effect, the way $\tau_0$ and $\tau_z$ enter
Eq.~(\ref{eq:npGtunn}) is switched as compared with Eq.~(\ref{eq:nnGtunn}),
and the same holds for $\tau_\|$ and $\tau_\perp$. The reason for this is the fact that
the pseudo-spin of eigenstates at a given wave vector in the conduction band is opposite
to the eigenstate with the same wave vector in the valence band. For conserved
pseudo-spin and equal densities, the obtained result
\begin{equation}\label{eq:npGsp}
\frac{G^{(\text{slg})}_{n\leftrightarrow p}(0)}{G_0} = \frac{Q}{\sqrt{4 \bar
k_{\text{F}}^2 - Q^2}}\,\, \Theta(2\bar k_{\text{F}} - Q)
\end{equation}
\end{subequations}
coincides with the one found for tunneling between parallel surfaces of a topological
insulator~\cite{per12}.

\begin{figure*}
\includegraphics[width=0.9\columnwidth]{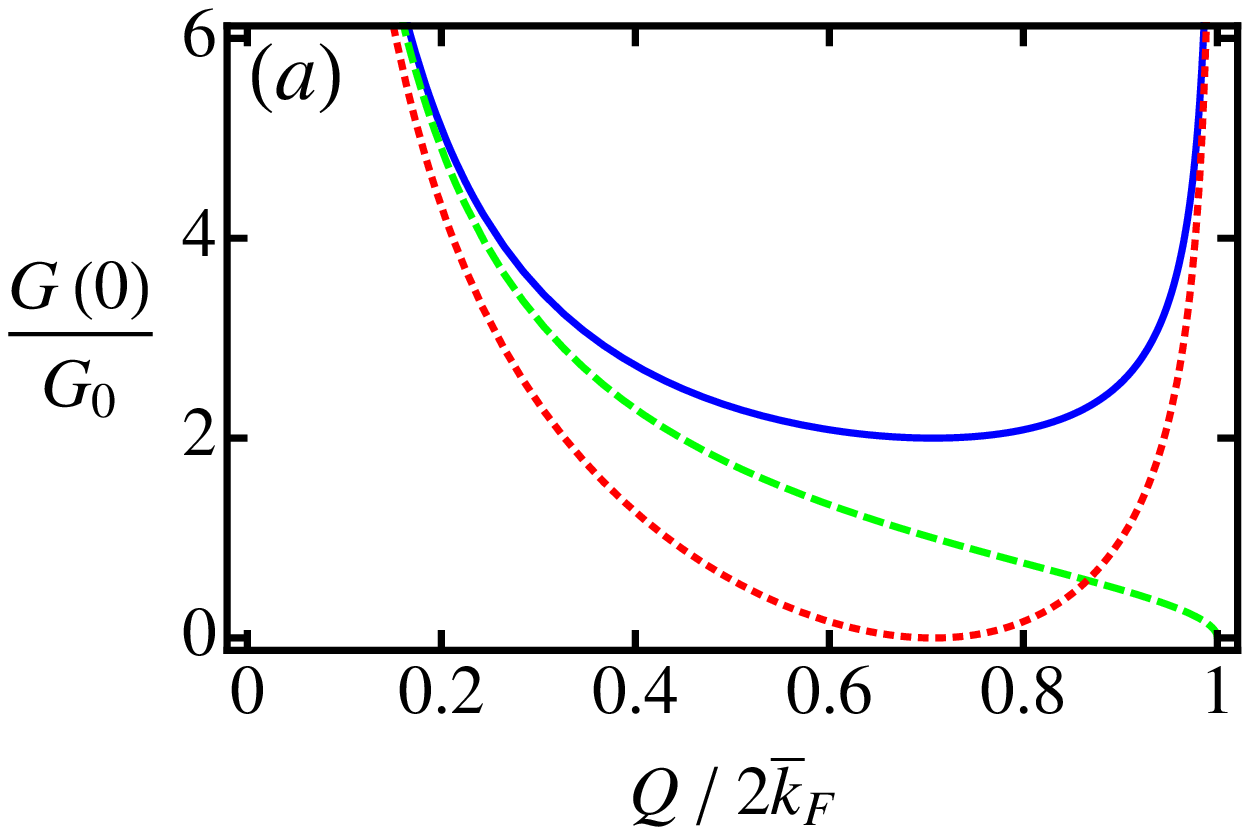}\hspace{1cm}
\includegraphics[width=0.9\columnwidth]{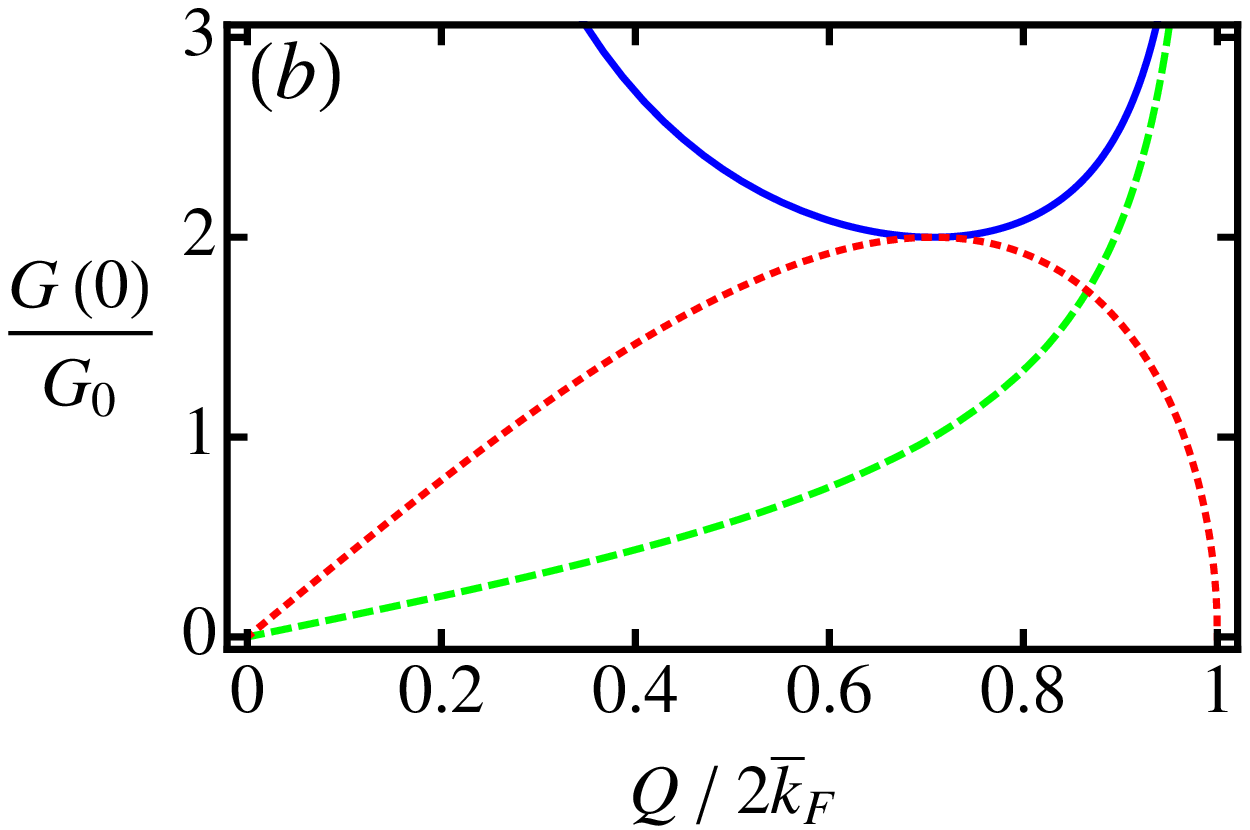}
\caption{\label{fig:condplots}
Linear magneto-tunneling conductances between parallel ordinary 2D electron
systems (blue solid curves), single-layer graphene sheets (green dashed curves),
and bilayer-graphene sheets (red dotted curves) through a pseudo-spin-conserving
barrier. $Q=d/\ell_{B_\|}^2$ is the wave-vector boost induced by a magnetic field
of magnitude $B_\|$ parallel to the two 2D systems, with $d$ denoting the latters'
vertical separation and $\ell_{B_\|}=\sqrt{\hbar/|e B_\||}$ the magnetic length.
Panel (a) [(b)] shows results for the case when tunneling occurs between two
\textit{n}-type layers [between an \textit{n}-type and a \textit{p}-type layer] with
equal density. The pseudo-spin structure of chiral electron states in single-layer and
bilayer graphene is the origin of the strongly modified magnetic-field dependences
of the tunneling conductance as compared with the ordinary 2D-electron case.}
\end{figure*}

Electrons in a graphene bilayer~\cite{mcc06} have energy dispersion
$\varepsilon_{\gamma,\kk,\sigma}^{(\text{blg})} = \sigma\, \hbar^2 k^2/(2 M)$
and pseudo-spin states
\begin{equation}
\ket{\sigma}_{\vek{K},\kk}^{(\text{blg})} = \frac{1}{\sqrt{2}} \left(\begin{array}{c}
\ee^{i \theta_\kk} \\ -\sigma \ee^{-i \theta_\kk} \end{array} \right) \, ,
\, \ket{\sigma}_{\vek{K'},\kk}^{(\text{blg})} = \frac{1}{\sqrt{2}} \left(
\begin{array}{c} \ee^{-i \theta_\kk} \\ -\sigma \ee^{i \theta_\kk}
\end{array} \right) .
\end{equation}
The full analytical expressions for the magneto-tunneling conductance between
parallel graphene bilayers are quite cumbersome and therefore given in
Eqs.~(\ref{eq:BLGlin}) of Appendix~\ref{sec:appA}. For equal densities in both
systems and a pseudo-spin-conserving barrier, we find
\begin{subequations}
\begin{eqnarray}
\frac{G^{(\text{blg})}_{n\leftrightarrow n}(0)}{G_0} &=& \frac{\left( 2 \bar k_{F}^2 -
Q^2 \right)^2}{\bar k_{\text{F}}^2 \,Q \, \sqrt{4 \bar k_{F}^2 - Q^2}} \,\, \Theta(2
\bar k_{\text{F}} - Q) \quad , \\
\frac{G^{(\text{blg})}_{n\leftrightarrow p}(0)}{G_0} &=& \frac{Q\, \sqrt{4 \bar
k_{\text{F}}^2 - Q^2}}{\bar k_{\text{F}}^2} \,\, \Theta(2 \bar k_{\text{F}} - Q)
\quad .
\end{eqnarray}
\end{subequations}

Figure~\ref{fig:condplots} illustrates the drastically different features
in the linear magneto-tunneling characteristics for single-layer and bilayer
graphene as compared with the ordinary 2D-electron-gas case. For
definiteness, the case of a pseudo-spin-conserving barrier is exhibited. The
overall suppression of tunneling transport between chiral 2D systems is a
consequence of the, in general, misaligned pseudo-spin polarizations of
states where the two systems' Fermi surfaces intersect. (See
Fig.~\ref{fig:fermicircles}.) In particular, pseudo-spin orthogonality leads to
the vanishing of $G(0)$ in a system of two \textit{n}-type single layers (bilayers)
of graphene when $Q = 2\bar k_{\text{F}}$ ($Q = \sqrt{2}\,\bar k_{\text{F}}$).
The fact that the pseudo-spin for eigenstates with opposite sign of the energy
is reversed results in the interchange of minima and maxima/divergences in
$G(0)$ for tunneling between an \textit{n}-type and a \textit{p}-type layer as
compared with the case of tunneling between two \textit{n}-type layers. The
magneto-tunneling conductance of the ordinary 2D electron system is reached
whenever the pseudo-spins of tunneling states are aligned, e.g., for $Q =
\sqrt{2}\,\bar k_{\text{F}}$ in tunneling between an \textit{n}-type and a
\textit{p}-type graphene bilayer. The possibility to have pseudo-spin flipped in
a tunneling process enables an even richer structure for tunneling transport,
which is captured for the completely general case by the formulae given in
Eqs.~(\ref{eq:nnGtunn}) and (\ref{eq:npGtunn}) [Eqs.~(\ref{eq:BLGlin})] for the
single-layer [bilayer] graphene case.

\begin{figure*}
\includegraphics[height=4.5cm]{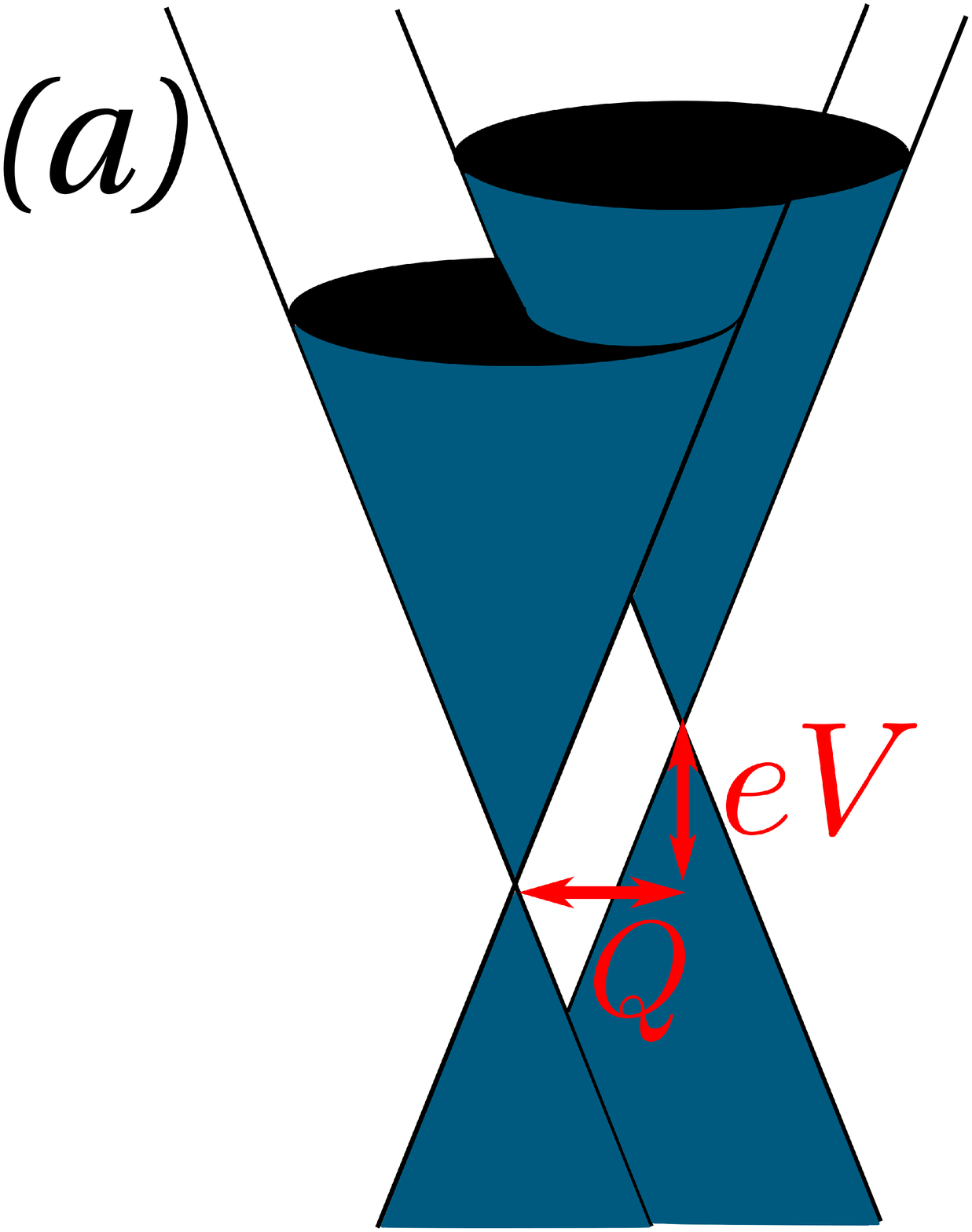}\hfill
\includegraphics[height=5cm]{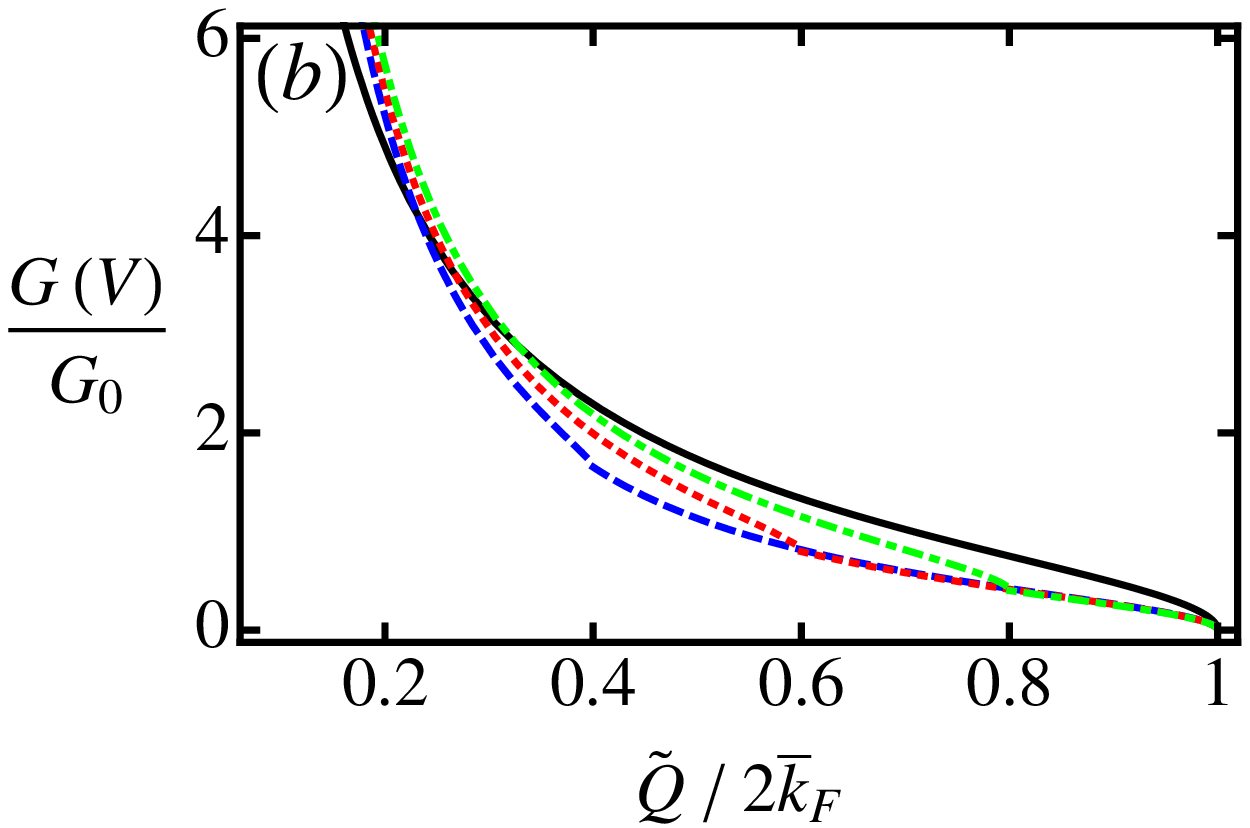}\hfill
\includegraphics[height=5cm]{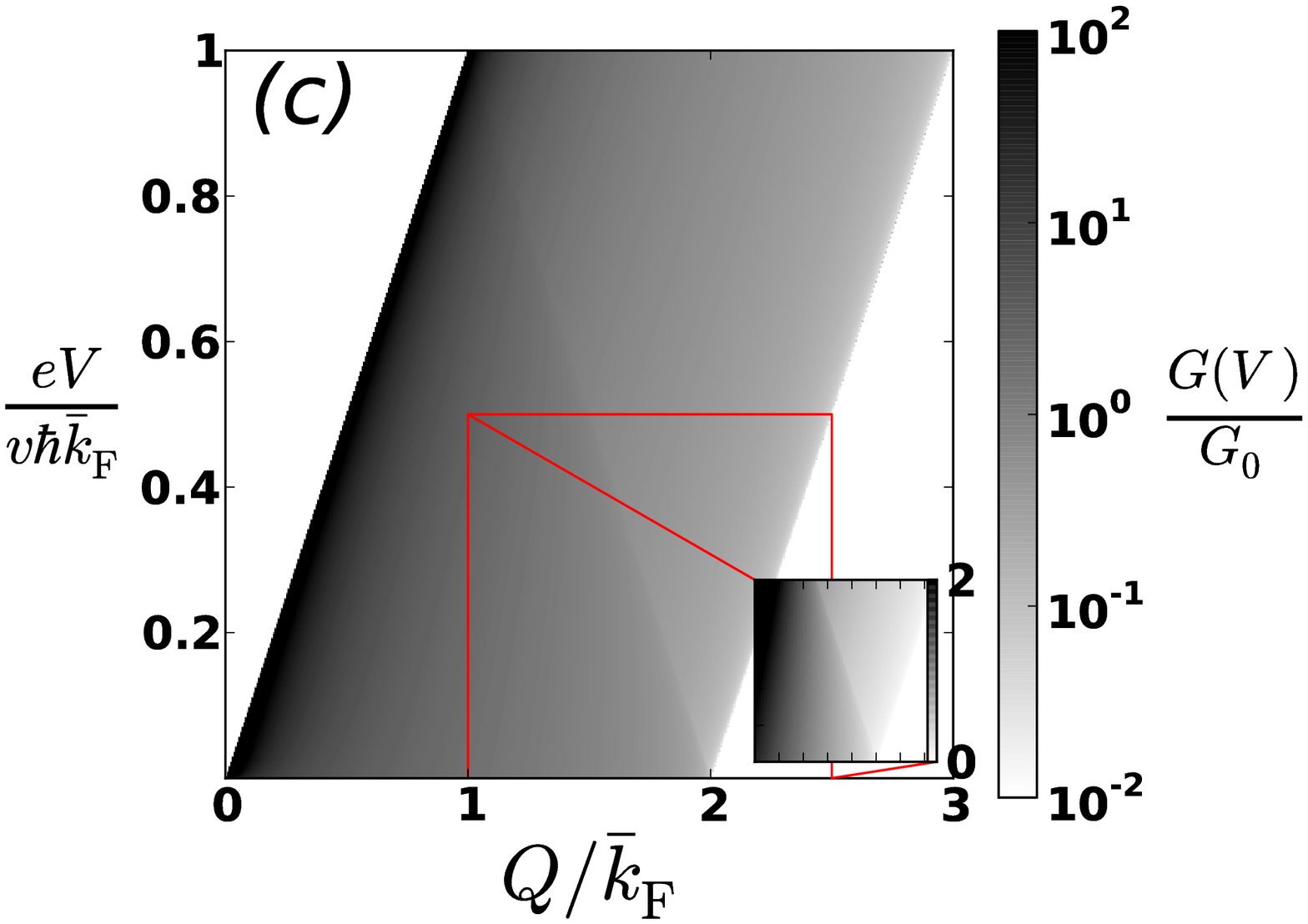}
\caption{\label{fig:finiteBias}
Magneto-tunneling between parallel single-layer graphene sheets at finite bias.
(a)~The two systems' Dirac-cone dispersions are shifted with respect to each other
by $Q$ ($eV$) in wave-vector (energy) direction due to an applied in-plane magnetic
field (bias voltage). Tunneling is possible for states where the conical surfaces intersect,
if the state is occupied in one layer and unoccupied in the other. (b)~Differential
tunneling conductance $G(V)$ for the case when both layers have equal \textit{n}-type
carrier density, plotted as a function of $\tilde Q = Q - eV/(\hbar v)$ for $e V=0$ (black
solid curve), $0.2\, \hbar v \bar k_{\text{F}}$ (green dot-dashed curve), $0.4\, \hbar v
\bar k_{\text{F}}$ (red dotted curve), and $0.6\, \hbar v \bar k_{\text{F}}$ (blue dashed
curve). $G(V)$ diverges at $\tilde Q=0$ and exhibits square-root-like features at
$\tilde Q = 2\bar k_{\text{F}} - 2 e V/(\hbar v)$ and $\tilde Q = 2\bar k_{\text{F}}$.
(c)~Logarithmic gray-scale plot of the differential tunneling conductance. The
divergence at $eV = \hbar v Q$ and the conical feature with apex at $Q=2 \bar
k_{\text{F}}$ constitute direct measures for the energy dispersion of charge carriers
in single-layer graphene.}
\end{figure*}

In contrast to single-layer and bilayer graphene, which are conductors, a
single layer of MoS$_2$ is a semiconducting 2D material. The electronic
dispersion is~\cite{xia12,kor13} $\varepsilon_{\gamma,\kk,\sigma}^{(\text{mos})}
=\sigma\, \hbar v \sqrt{k^2 + k_\Delta^2}$, with constant $k_\Delta>0$, and
using the abbreviation $\zeta_k = k_\Delta/\sqrt{k^2 + k_\Delta^2}$, the
pseudo-spin states can be expressed as
\begin{eqnarray}
&& \ket{\sigma}_{\vek{K},\kk}^{(\text{mos})} = \left( \begin{array}{c}
\sqrt{\frac{1+ \sigma \zeta_k}{2}} \, \ee^{-i \frac{\theta_\kk}{2}}  \\ \sigma
\sqrt{\frac{1- \sigma \zeta_k}{2}} \, \ee^{i \frac{\theta_\kk}{2}} \end{array}
\right) \,\, , \nonumber \\
&& \hspace{2.5cm} \ket{\sigma}_{\vek{K'},\kk}^{(\text{mos})} =
\left( \begin{array}{c} \sqrt{\frac{1+ \sigma \zeta_k}{2}} \, \ee^{- i \frac{\pi -
\theta_\kk}{2}} \\ \sigma \sqrt{\frac{1- \sigma \zeta_k}{2}} \, \ee^{i \frac{\pi -
\theta_\kk}{2}}\end{array} \right) . \quad
\end{eqnarray}
The most general expression for the linear magneto-tunneling conductance
between parallel single-layer MoS$_2$ systems is very complicated, and even
the result for a pseudo-spin-conserving barrier is so long that it has been relegated
to Appendix~\ref{sec:appA} [see Eqs.~(\ref{eq:MoS2lin})]. If in addition the densities
in both layers are equal, we find for the two doping configurations
\begin{subequations}
\begin{eqnarray}
\frac{G^{(\text{mos})}_{n\leftrightarrow n}(0)}{G_0} &=& \left( \frac{\sqrt{4 \bar
k_{\text{F}}^2 - Q^2}}{Q} + \frac{\zeta_{\bar k_{\text{F}}}^2 Q}{\sqrt{4 \bar
k_{\text{F}}^2 - Q^2}} \right) \nonumber \\[0.2cm] && \hspace{2.5cm}\times\,
\Theta(2 \bar k_{\text{F}} - Q) \, , \\
\frac{G^{(\text{mos})}_{n\leftrightarrow p}(0)}{G_0} &=& \frac{\big( 1 - \zeta_{\bar
k_{\text{F}}}^2 \big) Q}{\sqrt{4 \bar k_{\text{F}}^2 - Q^2}} \,\, \Theta(2 \bar
k_{\text{F}} - Q) \quad .
\end{eqnarray}
\end{subequations}
As expected, the behavior of MoS$_2$ in the limit $\zeta_k\to 0$ is the
same as that exhibited by single-layer graphene. See Eqs.~(\ref{eq:nnGsp})
and (\ref{eq:npGsp}). For $\zeta_k\to 1$, $G^{(\text{mos})}_{n\leftrightarrow n}(0)$
recovers the result (\ref{eq:ord2D}) found for an ordinary 2D electron system,
whereas pseudo-spin conservation causes $G^{(\text{mos})}_{n\leftrightarrow p}
(0)$ to vanish. 

\section{Magneto-tunneling at finite bias
\label{sec:finBiasG}}

Application of the general formula (\ref{eq:tuncurr}) to momentum-resolved tunneling
between parallel 2D electron systems in the zero-temperature limit and without
disorder yields the general expression
\begin{equation}\label{eq:nonlinI}
I(V) = \frac{1}{e} \int_{\varepsilon_{\text{F}}}^{\varepsilon_{\text{F}} + e V} \,
d\varepsilon \,\,\, \tilde G(\varepsilon, V) \quad .
\end{equation}
Here $\varepsilon_{\text{F}}$ is the Fermi energy of the 2D system whose subband
edge (or neutrality point) is taken as the zero of energy. The function
$\tilde G(\varepsilon, V)$ corresponds to the linear tunneling conductance between
the two 2D systems when the chemical potential is equal to $\varepsilon$ and $eV$
has been added to the zero-bias subband-edge splitting.

For illustration of the general principle, we focus here on the special case of
pseudo-spin-conserving tunneling between two \textit{n}-type single-layer graphene
sheets with equal carrier densities. It is then straightforward to find
\begin{eqnarray}
\tilde G(\varepsilon, V) &=& G_0 \,\, \sqrt{\frac{( 2 \varepsilon - e V)^2 - (\hbar v Q
)^2}{(\hbar v Q)^2 - (eV)^2}} \nonumber  \\[0.2cm] && \hspace{0.1cm} \times \,
\Theta \left( |2 \varepsilon - e V|  - \hbar v Q\right) \Theta(\hbar v Q - |eV|) \quad
\end{eqnarray}
by specializing the expression (\ref{eq:nnGtunn}) to the situation with $\tau_{\perp,
\|,z} = 0$ as well as making the substitutions $2 \bar k_{\text{F}} \to (2\varepsilon -
eV)/(\hbar v)$ and $\Delta \to eV/(\hbar v)$. Calculation of the current using
(\ref{eq:nonlinI}) and taking the derivative with respect to $V$ yields the differential
magneto-tunneling conductance $G(V)$ shown in Fig.~\ref{fig:finiteBias}. It switches
on with a divergence when $Q = |e V|/(\hbar v)$ and also exhibits features for $Q =
2\bar k_{\text{F}} \pm |e V|/(\hbar v)$, which mirror the characteristic switching-off
behavior seen in the linear magneto-tunneling conductance between graphene
layers at $Q = 2\bar k_{\text{F}}$ [see the green dashed curve in
Fig.~\ref{fig:condplots}(a)].

Characteristic features in the differential tunneling conductance between ordinary
(non-chiral) 2D electron systems have been shown to provide a direct image of
the electronic dispersion relation~\cite{rai95,lyo06,bie06}. The same applies to
magneto-tunneling at finite bias between chiral 2D electron systems, except that
the type of feature (e.g., divergence, or vanishing) of the differential conductance
associated with a dispersion branch is determined by pseudo-spin overlaps. For
example, in contrast to the ordinary 2D-electron case where the individual systems'
dispersions are imaged by peaks in the $Q$-dependence of $G(V)$, certain
dispersion branches from single-layer graphene sheets are drawn by a square-root-like
turning-off behavior in magneto-tunneling transport. See Fig.~\ref{fig:finiteBias}. 

\begin{figure*}
\includegraphics[height=4.9cm]{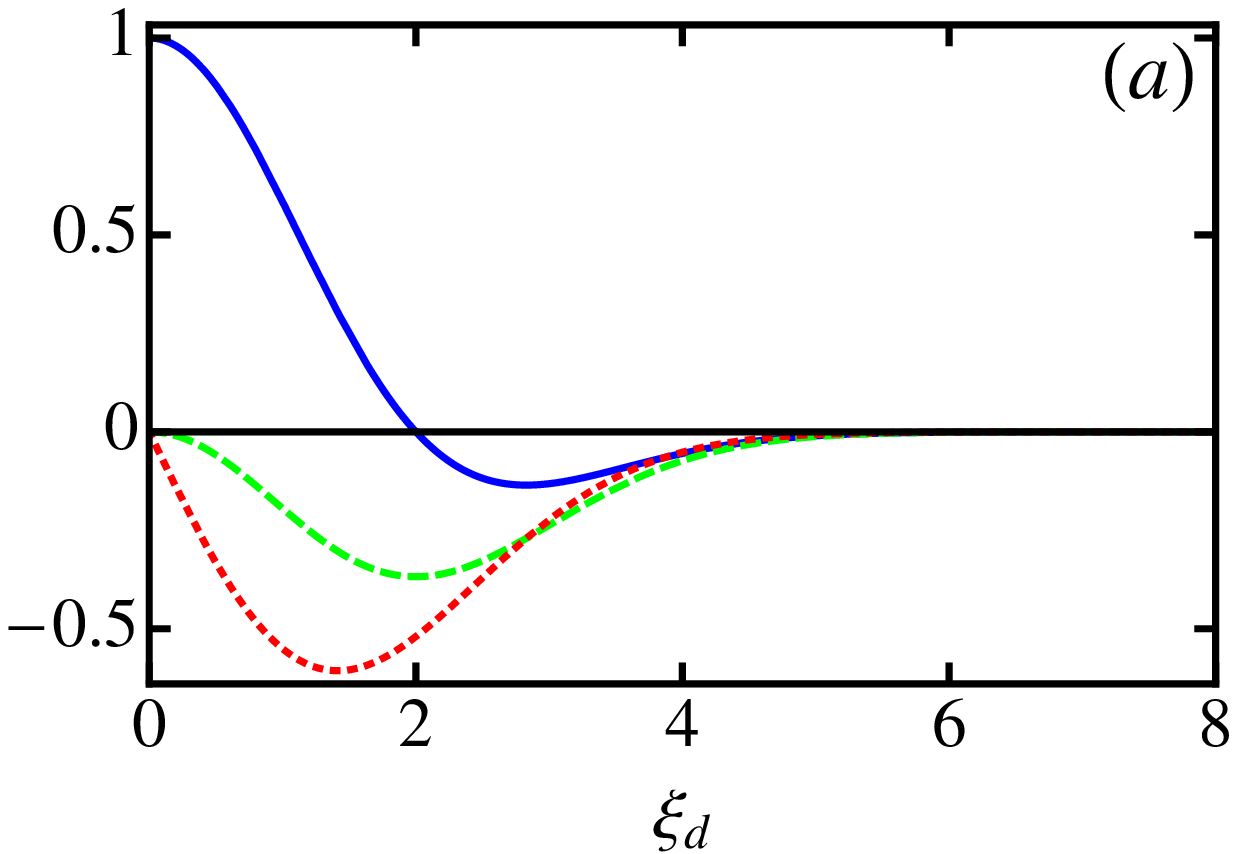}\hspace{1cm}
\includegraphics[height=4.9cm]{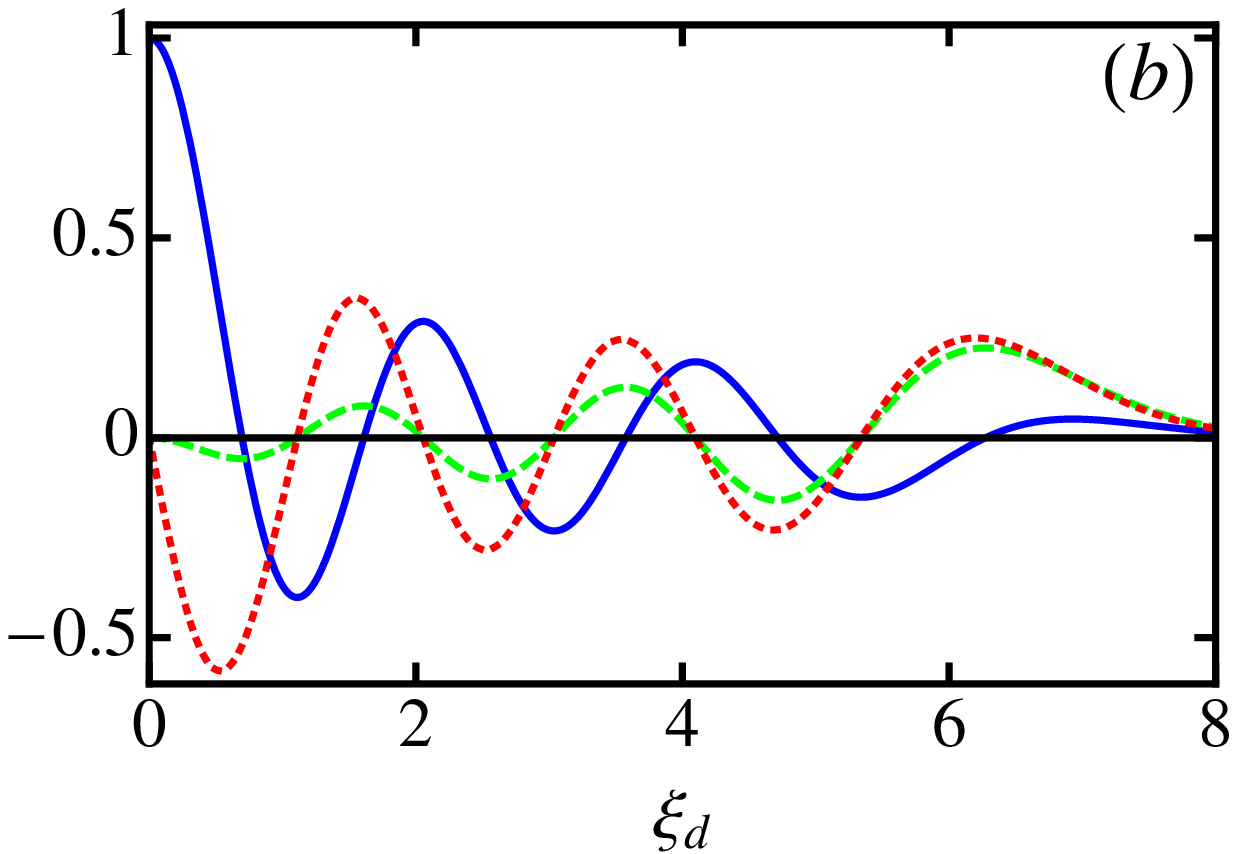}
\caption{\label{fig:LLtun}
Form factors for tunneling between graphene layers spaced at distance $d$ in a
tilted magnetic field $\vek{B}=(\vek{B}_\|, B_\perp)$, plotted as a function of the
parameter $\xi_d = (d/\ell_{B_\perp})(B_\| / B_\perp)$. See Eq.~(\ref{eq:LLcond}).
Panel~(a) [(b)] shows $\mathcal{F}^{(+)}_\nu(\xi_d)$ (blue solid curve),
$\mathcal{F}^{(-)}_\nu(\xi_d)$ (green dashed curve), and $\mathcal{F}^{(\perp)}_\nu
(\xi_d)$ (red dotted curve) for $\nu=1$ [$\nu=6$]. Note the limiting behavior for
$\xi_d\to 0$ and the oscillatory behavior for cases with $\nu>1$.}
\end{figure*}

\section{Magneto-tunneling between Landau-quantized graphene layers
\label{sec:LandQuant}}

The linear tunneling conductance between two chiral 2D electron systems in
the presence of a non-vanishing \emph{perpendicular\/} magnetic-field
component can be found by straightforward application of the general formula
(\ref{eq:tuncond}). Here we discuss in greater detail the case of parallel single
layers of graphene. Using the form (\ref{eq:kresolvedT}) for the tunneling matrix
and Landau-level eigenstates and -energies for graphene~\cite{cas09,goe11}, we
find analytic results presented in detail in Appendix~\ref{sec:appB}. As previously,
we focus on the zero-temperature limit and a system without disorder. (Both
of these assumptions can be relaxed straightforwardly in principle, resulting in
the usual smoothening of resonant features.) To illustrate the effects arising from
pseudo-spin dependence, we consider $G(0)$ for the special case when both
layers have equal density:
\begin{widetext}
\begin{subequations}\label{eq:LLcond}
\begin{eqnarray}
G^{(\mathrm{LLg})}_{n\leftrightarrow n}(0) &=& \frac{g_{\text{s}} g_{\text{v}}
e^2}{\hbar}\, \frac{A}{\hbar^2 v^2}\, \nu_{\text{F}} \sum_{\nu_1, \nu_2 = 1}^\infty
\delta \left(\nu_{\text{F}} - \nu_1\right)\delta \left( \nu_{\text{F}} - \nu_2 \right)
\left[ \left| \tau_0\, \mathcal{F}^{(+)}_{\nu_1}(\xi_d) + \tau_\perp\,
\mathcal{F}^{(\perp)}_{\nu_1}(\xi_d) \right|^2 + \left| \tau_z\,
\mathcal{F}^{(-)}_{\nu_1}(\xi_d)\right|^2 \right] \,\, , \\
G^{(\mathrm{LLg})}_{n\leftrightarrow p}(0) &=& \frac{g_{\text{s}} g_{\text{v}}
e^2}{\hbar}\, \frac{A}{\hbar^2 v^2}\, \nu_{\text{F}} \sum_{\nu_1, \nu_2 = 1}^\infty
\delta \left(\nu_{\text{F}} - \nu_1\right)\delta \left( \nu_{\text{F}} - \nu_2 \right)
\left[ \left| \tau_0\, \mathcal{F}^{(-)}_{\nu_1}(\xi_d) \right|^2 + \left| \tau_\|\,
\mathcal{F}^{(\perp)}_{\nu_1}(\xi_d) + \tau_z\, \mathcal{F}^{(+)}_{\nu_1} (\xi_d)
\right|^2 \right] \,\, .
\end{eqnarray}
\end{subequations}
\end{widetext}
Here $\nu_{\text{F}}$ is the Landau level at the Fermi energy, and the dependence
on the in-plane magnetic-field component is governed by form factors
$\mathcal{F}^{(\pm,\perp)}(\xi_d)$ through the parameter $\xi_d = (d/\ell_{B_\perp})
(B_\| / B_\perp)$. See Fig.~\ref{fig:LLtun} and explicit mathematical expressions
given in Appendix~\ref{sec:appB}. The oscillatory behavior as a function of $B_\|$
exhibited by the form factors originates from conservation of canonical momentum,
which restricts tunneling to Landau-level eigenstates with $B_\|$-dependent
displacement of their guiding-center locations~\cite{rai97,lyo98}. The linear conductance
oscillates also as a function of $B_\perp$ because of the Landau-quantization of
eigenenergies in 2D electron systems~\cite{smo89,rai97,lyo98}.

The chiral nature of charge carriers in graphene is manifested in a number of
differences with respect to the case of an ordinary 2D electron system that was
studied, e.g., in Refs.~\onlinecite{rai97,lyo98}. Instead of just one form factor that
depends on the in-plane field component~\cite{rai97,lyo98}, there are four different
form factors in the graphene case, each associated with an independent contribution
$\propto \tau_j$ to the, in general, pseudo-spin-dependent tunneling matrix. If both
graphene layers have equal density, one such form factor vanishes identically. In
the limit $B_\|\to0$, only one form factor remains finite, and the linear tunneling
conductance becomes proportional to $|\tau_0|^2$ ($|\tau_z|^2$) for a system
with two \textit{n}-type layers (one \textit{n}-type and one \textit{p}-type layer). Thus
linear tunneling transport between Landau-quantized graphene layers enables the
direct extraction of pseudo-spin-dependent tunneling matrix elements. This feature
will aid in our proposed scheme to extract quantitative information about the pseudo-spin
properties of the vertical heterostructure, which is described in the following Section.

\section{How to extract the pseudo-spin structure of the tunneling matrix
\label{sec:ExtractT}}

Our above considerations have shown how tunneling transport between chiral 2D
electron systems is strongly dependent on the pseudo-spin structure of the tunnel
coupling. As pseudo-spin is related to sub-lattice position, a full parametric study
of the tunneling conductance could be employed to yield information about
morphological details of the vertical heterostructure. While any type of chiral 2D
system lends itself to such an investigation, we describe below an approach that
works for two parallel single layers of graphene. 

Measurement of the magneto-tunneling conductance between two graphene layers
as a function of the externally adjustable parameters $Q$, $\bar k_{\text{F}}$
and $\Delta$ makes it possible to extract information about the tunneling matrix
$\tau$ given in Eq.~(\ref{eq:PauliTun}). This can be done because, according to
Eq.~(\ref{eq:nnGtunn}), the function
\begin{subequations}
\begin{equation}
F(Q, \bar k_{\text{F}}, \Delta) = \frac{2\pi\hbar G(0)}{g_{\text{s}} g_{\text{v}} e^2}
\, Q^2 \sqrt{(4 \bar k_{\text{F}}^2 - Q^2)(Q^2 - \Delta^2)}
\end{equation}
is a homogeneous polynomial of its arguments,
\begin{equation}\label{eq:polynom}
F(Q, \bar k_{\text{F}}, \Delta) \equiv -c_1\, Q^4 + c_2\, Q^2 \bar k_{\text{F}}^2
- c_3\, Q^2 \Delta^2 + c_4\, \bar k_{\text{F}}^2 \Delta^2 \, ,
\end{equation}
with coefficients
\begin{eqnarray}
c_1 &=& \frac{A \left( |\tau_0|^2 - |\tau_z|^2 \right)}{\hbar^2 v^2} \,\,\, , \,\,\,
c_2 = 4\, \frac{A \left( |\tau_0|^2 + |\tau_\||^2 \right)}{\hbar^2 v^2} \, ,
\nonumber \\
c_3 &=& \frac{A \left( |\tau_\perp|^2 + |\tau_z|^2 \right)}{\hbar^2 v^2} \,\, , \,\, 
c_4 = 4\, \frac{A \left( |\tau_\perp|^2 - |\tau_\||^2 \right)}{\hbar^2 v^2} \, . \quad
\end{eqnarray}
\end{subequations}
Performing fits of the obtained data to the polynomial form (\ref{eq:polynom})
yields the coefficients $c_j$. For example, a possible strategy could be to
start with measuring $G(0)$ as a function of $Q$ for equal densities in the layers
and using the form of $F(Q, \bar k_{\text{F}}, 0)$ to determine $c_1$ and $c_2$.
Fixing then a particular value of $\bar k_{\text{F}}$ and $\Delta\ne 0$, varying
only $Q$ and considering the combination $F(Q, \bar k_{\text{F}}, \Delta) + c_1
Q^2 - c_2 Q^2 \bar k_{\text{F}}^2$ will then enable extraction of $c_3$ and
$c_4$ from a fit to this quantity's $Q$ dependence. A first reality check for the
theory proposed here would be to demonstrate the relation $c_2 + c_4 = 4(c_1
+ c_3)$. 

The fact that the coefficients $c_j$ satisfy a linear relation means that we need
an additional independent measurement to determine the magnitudes of tunnel
matrix elements. Resonant tunneling transport in a quantizing perpendicular
magnetic field for equal densities between the layers can be used for this purpose.
Application of Eqs.~(\ref{eq:LLcond}) allows to extract the ratio of $|\tau_0|^2/
|\tau_z|^2$, assuming that the inelastic scattering time that broadens the tunneling
resonances is the same for \textit{n}-type and \textit{p}-type graphene layers. Then
all magnitudes of tunneling matrix elements can be determined in units of $\hbar^2
v^2/A$.

The freedom to change the in-plane field direction enables further information to be
extracted from magneto-tunneling measurements. A general expression for the
magnitudes of tunneling matrix elements can be given in terms of the azimuthal
angle $\theta_{\vek{B}_\|}\equiv\arctan(B_{\|, y}/B_{\|, x})$ of the in-plane magnetic
field,
\begin{subequations}
\begin{eqnarray}
|\tau_\perp(\theta_{\vek{B}_\|})|^2 &=& \frac{|\tau_x|^2 + |\tau_y|^2}{2} +
\frac{|\tau_x|^2 - |\tau_y|^2}{2} \cos(2\theta_{\vek{B}_\|}) \nonumber \\ && 
 \hspace{1.5cm} +\, \Re \{\tau_x \tau_y^\ast \}\sin(2\theta_{\vek{B}_\|}) \,\, ,\quad \\
|\tau_\|(\theta_{\vek{B}_\|}) |^2 &=& \frac{|\tau_x|^2 + |\tau_y|^2}{2} -
\frac{|\tau_x|^2 - |\tau_y|^2}{2} \cos(2\theta_{\vek{B}_\|}) \nonumber \\ && 
 \hspace{1.5cm} -\, \Re\{ \tau_x\tau_y^\ast \}\sin(2\theta_{\vek{B}_\|}) \,\, .
\end{eqnarray}
\end{subequations}
Thus the phase difference between the generally complex-valued matrix elements
$\tau_x$ and $\tau_y$ can be determined from the tunneling-matrix magnitudes
found for $\theta_{\vek{B}_\|}=0$ and $\theta_{\vek{B}_\|}=\pi/4$:
\begin{equation}
\arg(\tau_x\tau_y^\ast) = \arccos\left[\frac{|\tau_\perp(\frac{\pi}{4})|^2 - |\tau_\perp
(\frac{\pi}{4})|^2}{2|\tau_\|(0)||\tau_\perp(0)|} \right] \quad .
\end{equation}

\section{Discussion and Conclusions
\label{sec:concl}}

Experimental exploration of the magneto-tunneling characteristics discussed
above requires sufficiently large magnetic fields to shift the entire Fermi circle
in kinetic-wave-vector space. Specifically, the condition $|z_2 - z_1|\equiv d \ge
2\bar k_{\text{F}} \, \ell_{B_\parallel^\mathrm{(max)}}^2$ ensures that the full
range of fields over which tunneling occurs can be accessed. For the case of
equal density $n = g_{\text{s}} g_{\text{v}} \bar k_{\text{F}}^2 / (4\pi)$ in the two
layers, we find
\begin{equation}
B_\parallel^\mathrm{(max)} \ge \frac{2\pi\hbar}{e}\, \sqrt{\frac{4}{g_{\text{s}}
g_{\text{v}}}\frac{n}{\pi d^2}} \approx 20\,\mathrm{T}\times \frac{\sqrt{n\, [10^{10}
\, \mathrm{cm}^{-2}]}}{d\, [ \mathrm{nm} ] } \,\, .
\end{equation}
As encapsulation of graphene sheets was shown to enable ballistic transport
over $\mu$m-scale distances at low carrier densities~\cite{dea10,may11},
devices with $B_\parallel^\mathrm{(max)}$ within routinely reachable
limits should be accessible with current technology.

Inelastic scattering of 2D chiral quasi-particle excitations due to impurities,
coupling to phonons, or Coulomb interactions results in their finite lifetime
and concomitant broadening of resonant behavior in the magneto-tunneling
conductance~\cite{zhe93,lyo93,rai96,jun96,rai97,lyo98}. Such effects can be
straightforwardly included in the calculation based on Eq.~(\ref{eq:tuncond})
by using the appropriate form of the single-electron spectral function with life-time
broadening.

In conclusion, we have derived analytical expressions for the magneto-tunneling
conductance between parallel layers of graphene, bilayer graphene, and MoS$_2$
in the low-temperature limit and in the absence of interactions and disorder. The
constraints imposed by simultaneous energy and momentum conservation in
the tunneling processes result in characteristic dependencies on in-plane and
perpendicular-to-the-plane magnetic fields as well as the bias voltage. The
pseudo-spin properties and chirality of charge carriers in the vertically separated
layers strongly affect the magneto-tunneling transport features. Based on the
additional dependencies on the densities/Fermi wave vectors in each layer, it is
possible to determine the pseudo-spin structure of the tunnel barrier. Our work
can thus be used to study, and optimize the design of, vertical-tunneling structures
between novel two-dimensional (semi-)conductors.

\acknowledgments
We would like to thank M.~Governale for useful discussions and helpful
comments on the manuscript. LP gratefully acknowledges financial support
from a Victoria University Master's-by-thesis Scholarship.

\appendix

\begin{widetext}

\section{Linear magneto-tunneling conductance for bilayer graphene and
MoS$_2$
\label{sec:appA}}

The general expression for the magneto-tunneling conductance between two
\textit{n}-doped bilayer-graphene layers is found to be
\begin{subequations}\label{eq:BLGlin}
\begin{eqnarray}
\frac{G^{({\rm blg})}_{n\to n}(0)}{G_0} &=& \frac{\Theta(Q-\Delta)\Theta(2\bar
k_{\text{F}}-Q)}{\mathrm{Tr}[\tau^\dagger\tau]} \left\{ \frac{\left| \tau_0 \left(
4\bar k_{\text{F}}^2 + \Delta^2 - 2 Q^2 \right) Q^2 - \tau_\perp \left[ 8\bar
k_{\text{F}}^2 \, \Delta^2 - \left( 4\bar k_{\text{F}}^2 + \Delta^2 \right) Q^2
\right]\right|^2}{Q^4 \left( 4\bar k_{\text{F}}^2 - \Delta^2 \right) \sqrt{\left(
4 \bar k_{\text{F}}^2 - Q^2 \right) \left( Q^2 - \Delta^2 \right)}} \right.
\nonumber \\ && \hspace{6cm} \left. + \,\, \left| \tau_\| \, 8 \bar
k_{\text{F}}^2\, \Delta^2 - i \, \tau_z \, 2 Q^2 \right|^2 \, \frac{\sqrt{\left( 4
\bar k_{\text{F}}^2 - Q^2 \right) \left( Q^2 - \Delta^2 \right)}}{Q^4 \left( 4\bar
k_{\text{F}}^2 - \Delta^2 \right)}\right\} \, , \quad 
\end{eqnarray}
whereas the conductance between an \textit{n}-doped and a \textit{p}-doped
bilayer is given by 
\begin{eqnarray}
\frac{G^{({\rm blg})}_{n\to p}(0)}{G_0} &=& \frac{\Theta(Q-\Delta)\Theta(2\bar
k_{\text{F}}-Q)}{\mathrm{Tr}[\tau^\dagger\tau]} \left\{ \left| \tau_0\, 2 Q^2 +
\tau_\perp \, 8 \bar k_{\text{F}}^2 \, \Delta^2 \right|^2 \frac{\sqrt{\left( 4 \bar
k_{\text{F}}^2 - Q^2 \right) \left( Q^2 - \Delta^2 \right)}}{Q^4 \left( 4\bar
k_{\text{F}}^2 - \Delta^2 \right)} \right. \nonumber \\ && \hspace{4cm} \left.
+\,\, \frac{\left| \tau_\| \left[ 8\bar k_{\text{F}}^2 \, \Delta^2 - \left( 4\bar
k_{\text{F}}^2 + \Delta^2 \right) Q^2 \right] + i\, \tau_z \left( 4\bar
k_{\text{F}}^2 + \Delta^2 - 2 Q^2 \right) Q^2 \right|^2}{Q^4 \left( 4\bar
k_{\text{F}}^2 - \Delta^2 \right) \sqrt{\left( 4\bar k_{\text{F}}^2 - Q^2 \right)
\left( Q^2 - \Delta^2 \right)}} \right\} \, . \quad
\end{eqnarray}
\end{subequations}
Note that, unlike for tunneling between single-layer graphene sheets, the phase
of the tunneling matrix plays a role in determining the transport characteristics
for tunneling between two bilayer-graphene systems. Furthermore, the conductance
obtained for tunneling between two \textit{p}-type bilayers differs from that found for
two \textit{n}-type bilayers by an opposite sign in the terms involving $\tau_\perp$
and $\tau_z$.

To discuss magneto-tunneling transport between two parallel single layers of
MoS${}_2$, we restrict ourselves to the case of a pseudo-spin-conserving barrier
because the fully general formulae are quite cumbersome. We obtain
\begin{subequations}\label{eq:MoS2lin}
\begin{eqnarray}
\frac{G^{(\mathrm{mos})}_{n\to n}(0)}{G_0} &=& \frac{\Theta(Q - \Delta) \Theta(2
\bar k_{\text{F}} - Q)}{4} \left\{ \sqrt{\frac{4 \bar k_{\text{F}}^2 - Q^2}{Q^2 - 
\Delta^2}} \left[ \sqrt{\left( 1 + \zeta_{k_{\text{F}}^{(1)}} \right) \left( 1 +
\zeta_{k_{\text{F}}^{(2)}} \right)} + \sqrt{\left( 1 - \zeta_{k_{\text{F}}^{(1)}} \right)
\left( 1 - \zeta_{k_{\text{F}}^{(2)}} \right)} \right]^2 \right. \nonumber \\
&& \hspace{2cm} \left. +\,\, \sqrt{\frac{Q^2 - \Delta^2}{4 \bar k_{\text{F}}^2 - Q^2}}
\left[ \sqrt{\left( 1+ \zeta_{k_{\text{F}}^{(1)}} \right) \left( 1 + \zeta_{k_{\text{F}}^{(2)}}
\right)} - \sqrt{\left( 1 - \zeta_{k_{\text{F}}^{(1)}} \right) \left( 1 -
\zeta_{k_{\text{F}}^{(2)}}\right)} \right]^2 \right\}
\end{eqnarray}
for the case when both layers are \textit{n}-doped, whereas for tunneling between
an \textit{n}-doped and a \textit{p}-doped layer, the result
\begin{eqnarray}
\frac{G^{(\mathrm{mos})}_{n\to p}(0)}{G_0}  &=& \frac{\Theta(Q - \Delta) \Theta(2
\bar k_{\text{F}} - Q)}{4} \left\{ \sqrt{\frac{4 \bar k_{\text{F}}^2 - Q^2}{Q^2 - 
\Delta^2}} \left[ \sqrt{\left( 1 + \zeta_{k_{\text{F}}^{(1)}} \right) \left( 1 -
\zeta_{k_{\text{F}}^{(2)}} \right)} - \sqrt{\left( 1 - \zeta_{k_{\text{F}}^{(1)}} \right)
\left( 1 + \zeta_{k_{\text{F}}^{(2)}} \right)} \right]^2 \right. \nonumber \\
&& \hspace{2cm} \left. +\,\, \sqrt{\frac{Q^2 - \Delta^2}{4 \bar k_{\text{F}}^2 - Q^2}}
\left[ \sqrt{\left( 1+ \zeta_{k_{\text{F}}^{(1)}} \right) \left( 1 - \zeta_{k_{\text{F}}^{(2)}}
\right)} + \sqrt{\left( 1 - \zeta_{k_{\text{F}}^{(1)}} \right) \left( 1 +
\zeta_{k_{\text{F}}^{(2)}}\right)} \right]^2 \right\}
\end{eqnarray}
\end{subequations}
is found.

\section{Momentum-resolved tunneling between Landau-quantized graphene layers
in a tilted field
\label{sec:appB}}

Using the familiar Landau-level ladder operators defined by $a^\pm=\ell_{B_\perp}
\left( \Pi_x \pm i\, \Pi_y \right)/(\sqrt{2}\hbar)$, with kinetic momentum $\vek{\Pi} =
\vek{p} + e\,\vek{A}$ in terms of the magnetic vector potential $\vek{A}$, the
single-particle Hamiltonians for the $\vek{K}$ and $\vek{K^\prime}\equiv -\vek{K}$
valleys of graphene are given by~\cite{cas09,goe11}
\begin{equation}
{\mathcal H}_{\pm\vek{K}}(B_\perp) = \pm \sqrt{2}\, \frac{\hbar v}{\ell_{B_\perp}}
\begin{pmatrix} 0 & a^\mp \\ a^\pm &0 \end{pmatrix} \quad .
\end{equation}
For definiteness, we choose the Landau gauge $\vek{A}=(-y\, B_\perp + z\, B_\|, 0, 0)$,
where $z$ is the constant $\vek{\hat z}$ coordinate of charge carriers in the 2D layer.
The energy eigenvalues of ${\mathcal H}_{\pm\vek{K}}(B_\perp)$ are found to be
$\epsilon_{\sigma,\nu} = \sigma\, \hbar v \sqrt{2\nu}/\ell_{B_\perp}$, where $\nu= 0, 1,
\dots$, and the corresponding eigenstates in the $\vek{K}$ and $\vek{K^\prime}$ valleys
are~\cite{cas09,goe11}
\begin{subequations}
\begin{eqnarray}
\ket{\nu, \sigma, \kappa_x}_{\vek{K}} &=& \frac{1}{\sqrt{2}} \begin{pmatrix} \sigma
\ket{\nu -1, \kappa_x} \\ \ket{\nu, \kappa_x} \end{pmatrix} \,\, \mbox{for $\nu > 0$, and}
\,\, \ket{0,\kappa_x}_{\vek{K}} = \begin{pmatrix} 0 \\ \ket{0, \kappa_x} \end{pmatrix} \quad ,
\\[0.2cm]
\ket{\nu, \sigma, \kappa_x}_{\vek{K^\prime}} &=&\frac{1}{\sqrt{2}} \begin{pmatrix} \ket{\nu,
\kappa_x} \\ \sigma \ket{\nu - 1,\kappa_x} \end{pmatrix}\,\, \mbox{for $\nu > 0$, and} \,\,
\ket{0, \kappa_x}_{\vek{K^\prime}} = \begin{pmatrix} \ket{0, \kappa_x} \\ 0\end{pmatrix}
\quad .
\end{eqnarray}
\end{subequations}
Here the real-space Landau-level eigenstates satisfy $a^+a^- \ket{\nu, \kappa_x} = \nu
\ket{\nu, \kappa_x}$, with the quantum number $\kappa_x \equiv k_x + z/\ell_{B_\|}^2$
being related to the cyclotron-orbit guiding-center position in $y$ direction. In the following,
it will be useful to note the mathematical relation~\cite{hu92,lyo98}
\begin{equation}\label{eq:shiftOL}
\left\langle \nu, \kappa_x\, | \, \nu^\prime, \kappa_x^\prime \right\rangle =
\delta_{k_x, k_x^\prime}\,\, (-1)^{\nu_>-\nu_<} \, \left(\frac{\nu_<!}{\nu_>!}\right)^{\frac{1}{2}}
\left( \frac{\xi^2}{2}\right)^{\frac{\nu_>-\nu_<}{2}} \ee^{-\frac{\xi^2}{4}} \,\, L_{\nu_<}^{\nu_>
-\nu_<} \left( \frac{\xi^2}{2} \right) \quad ,
\end{equation}
where $\nu_{<(>)} = \mathrm{min (max)}\{\nu, \nu^\prime\}$, $\xi = (|z - z'|/\ell_{B_\perp})
(B_\| / B_\perp)$, and $L_n^{n'}(\cdot)$ is the generalized Laguerre polynomial.

Using the Landau-level eigenstates and eigenenergies for calculating the linear tunneling
conductance from Eq.~(\ref{eq:tuncond}), we find
\begin{eqnarray}\label{eq:tiltBcond}
G^{(\mathrm{LLg})} &=& \frac{g_{\text{s}} g_{\text{v}} e^2}{\hbar}\, \frac{A}{\hbar^2 v^2}\,
\sqrt{\nu_{\text{F}}^{(1)} \left(\nu_{\text{F}}^{(1)}+\Delta\nu_{\text{F}} \right)} \sum_{\nu_1,
\nu_2 = 1}^\infty \delta \left( \nu_{\text{F}}^{(1)} - \nu_1\right)\delta \left( \nu_{\text{F}}^{(1)}
+\Delta\nu_{\text{F}} - \nu_2 \right) \nonumber\\ && \hspace{5.5cm} \times \left[ \left|
\tau_0\, F^{(0)}_{\nu_1 \nu_2}(\xi_d) + \tau_x\, F^{(x)}_{\nu_1 \nu_2}(\xi_d) \right|^2 + \left|
\tau_y \, F^{(y)}_{\nu_1 \nu_2}(\xi_d) + \tau_z\, F^{(z)}_{\nu_1 \nu_2}(\xi_d)\right|^2 \right]
\,\, , \quad 
\end{eqnarray}
where we denote the Landau level at the Fermi energy in layer $j$ by $\nu_{\text{F}}^{(j)}$,
$\Delta\nu_{\text{F}} = \nu_{\text{F}}^{(2)} - \nu_{\text{F}}^{(1)}$, and $\xi_d \equiv \xi$ for
$|z-z^\prime|\to d$ where $d$ is the vertical separation between the two graphene layers.
Terms with $\nu_1=0$ or $\nu_2=0$ have been omitted from the sum on the r.h.s\ of
(\ref{eq:tiltBcond}) because theses have a vanishing prefactor. It should be noted that
such terms \textit{would\/}, however, contribute if our assumption of purely elastic scattering
were to be relaxed. The $F^{(j)}_{\nu_1 \nu_2}(\xi)$ are form factors describing the effect of
the in-plane magnetic field. For $\nu_> \ne \nu_<$, we find
\begin{subequations}\label{eq:formNE}
\begin{eqnarray}
F^{(0)}_{\nu_1 \nu_2}(\xi)&=&\frac{1}{2} \, \left(\frac{\nu_<!}{\nu_>!}\right)^{\frac{1}{2}}
\left( \frac{\xi^2}{2}\right)^{\frac{\nu_>-\nu_<}{2}} \ee^{-\frac{\xi^2}{4}}\left[  L_{\nu_<}^{\nu_>
-\nu_<} \left( \frac{\xi^2}{2} \right) \pm \sqrt{\frac{\nu_>}{\nu_<}} \,\, L_{\nu_<-1}^{\nu_>
-\nu_<} \left( \frac{\xi^2}{2} \right)\right] \,\, ,
\\[0.2cm]
F^{(x)}_{\nu_1 \nu_2}(\xi)&=&-\frac{1}{2} \, \left(\frac{\nu_<!}{\nu_>!}\right)^{\frac{1}{2}}
\left( \frac{\xi^2}{2}\right)^{\frac{\nu_> - \nu_< - 1}{2}} \ee^{-\frac{\xi^2}{4}}\left[ \sqrt{\nu_>}
\,\, L_{\nu_<}^{\nu_> -\nu_<-1} \left( \frac{\xi^2}{2} \right) \pm \frac{\xi^2}{2 \sqrt{\nu_<}} \,\,
L_{\nu_<-1}^{\nu_> -\nu_<+1} \left( \frac{\xi^2}{2}\right) \right] \,\, , \\[0.2cm]
F^{(y)}_{\nu_1 \nu_2}(\xi)&=&-\frac{i}{2} \, \left(\frac{\nu_<!}{\nu_>!}\right)^{\frac{1}{2}}
\left( \frac{\xi^2}{2}\right)^{\frac{\nu_>-\nu_< -1}{2}} \ee^{-\frac{\xi^2}{4}}\left[ \sqrt{\nu_>}
\,\, L_{\nu_<}^{\nu_> -\nu_<-1} \left( \frac{\xi^2}{2} \right) \mp \frac{\xi^2}{2 \sqrt{\nu_<}} \,\,
L_{\nu_<-1}^{\nu_> -\nu_<+1} \left( \frac{\xi^2}{2}\right) \right] \,\, ,
\\[0.2cm]
F^{(z)}_{\nu_1 \nu_2}(\xi)&=&\frac{1}{2} \, \left(\frac{\nu_<!}{\nu_>!}\right)^{\frac{1}{2}}
\left( \frac{\xi^2}{2}\right)^{\frac{\nu_>-\nu_<}{2}} \ee^{-\frac{\xi^2}{4}}\left[  L_{\nu_<}^{\nu_>
-\nu_<} \left( \frac{\xi^2}{2} \right) \mp \sqrt{\frac{\nu_>}{\nu_<}}\,\, L_{\nu_<-1}^{\nu_>
-\nu_<} \left( \frac{\xi^2}{2} \right)\right] \,\, ,
\end{eqnarray}
\end{subequations}
\end{widetext}
where the upper (lower) sign of terms applies to tunneling between two \textit{n}-type layers
(an \textit{n}-type and a \textit{p}-type layer). When $\nu_1=\nu_2 \equiv \nu$, we have
\begin{subequations}\label{eq:formEQ}
\begin{eqnarray}
\left. F^{(0)}_{\nu \nu}(\xi) \right|_{n\to n}&\equiv& \left. F^{(z)}_{\nu \nu}(\xi) \right|_{n\to p}
= \mathcal{F}_\nu^{(+)}(\xi) \,\, , \\
\left. F^{(x)}_{\nu \nu}(\xi) \right|_{n\to n}&\equiv& \left. i\, F^{(y)}_{\nu \nu}(\xi) \right|_{n\to p}
= \mathcal{F}_\nu^{(\perp)}(\xi)  \,\, , \\
\left. F^{(y)}_{\nu \nu}(\xi)\right|_{n\to n}&\equiv& \left. F^{(x)}_{\nu \nu}(\xi) \right|_{n\to p}
=0 \,\, , \\
\left. F^{(z)}_{\nu \nu}(\xi) \right|_{n\to n}&\equiv& \left. F^{(0)}_{\nu \nu}(\xi) \right|_{n\to p}
= \mathcal{F}_\nu^{(-)}(\xi) \,\, ,
\end{eqnarray}
\end{subequations}
with the definitions
\begin{subequations}
\begin{eqnarray}
\mathcal{F}_\nu^{(\pm)}(\xi) &=& \frac{1}{2} \ee^{-\frac{\xi^2}{4}}\left[  L_{\nu}^{0} \left(
\frac{\xi^2}{2} \right) \pm L_{\nu-1}^{0} \left( \frac{\xi^2}{2} \right)\right] \,\, , \quad \\[0.2cm]
\mathcal{F}_\nu^{(\perp)}(\xi) &=& -\ee^{-\frac{\xi^2}{4}}\sqrt{\frac{\xi^2}{2\nu}} L_{\nu-1}^{1}
\left( \frac{\xi^2}{2} \right)  \,\, .
\end{eqnarray}
\end{subequations}

In the $B_\|=0$ limit (i.e., for $\xi\to 0$), the form factors restrict tunneling to occur between
 the same or adjacent Landau levels, depending on the pseudo-spin structure of the tunneling
matrix. 

%

\end{document}